\documentclass[onecolumn,showpacs,preprintnumbers,amsmath,amssymb,11pt]{revtex4}

\usepackage{epsf}
\usepackage{graphicx}  % Include figure files
\usepackage{dcolumn}   % Align table columns on decimal point
\usepackage{bm}        % bold math

\newcommand{\be}{\begin{equation}}
\newcommand{\en}{\end{equation}}
\newcommand{\bea}{\begin{eqnarray}}
\newcommand{\ena}{\end{eqnarray}}
\begin{document}

\preprint{}

\title{Semi-Holographic Universe }

\author{Hongsheng Zhang\footnote{Electronic address: hongsheng@kasi.re.kr} }
\affiliation{Shanghai United Center for Astrophysics (SUCA),
 Shanghai Normal University, 100 Guilin Road, Shanghai 200234,
 P.R.China}

 \author{Xin-Zhou Li \footnote{Electronic address: kychz@shnu.edu.cn} }
 \affiliation{Shanghai United Center for Astrophysics (SUCA), Shanghai Normal
University, 100 Guilin Road, Shanghai 200234, P.R.China}

 \author{Hyerim Noh\footnote{Electronic address: hr@kasi.re.kr} }
 \affiliation{
 Korea Astronomy and Space Science Institute,
  Daejeon 305-348, Korea }
\begin{abstract}
   By assuming  that a dark component (dark energy) in the universe strictly obeys
   the holographic principle, that is, its entropy is one fourth of
   the apparent horizon, we find that the existence of the other dark
   component (dark matter)  is compulsory, as a compensation of dark energy,
     based on the first law of
   thermodynamics. By using the method of dynamical system analysis, we find that
    there exists a stable dark energy-dark matter scaling solution at late
    time, which is helpful to solve the coincidence problem.
      For reasonable parameters, the deceleration parameter
   is well consistent with current observations.

\end{abstract}


\maketitle

\section{Introduction}

 The existence of dark energy is one of the most significant cosmological discoveries
 over the last century \cite{acce}.  Various models of dark energy have been
proposed, such as a small positive
 cosmological constant, quintessence, k-essence, phantom, holographic dark energy,
 etc., see \cite{review} for a
 recent review. However, although fundamental for
our understanding of the universe, its nature, especially in the
 theoretical aspect, remains a completely open question nowadays.

      The holographic principle is a very important idea in
      high energy physics and gets more and more attentions in different branches of physics. t' Hooft proposed
      the first version of the holographic principle and named it \cite{thsu}.
      In this version, the holographic principle declares that the true laws of inside any surface are actually a description
      of  how its image evolves on that surface. He guesses that the horizon of a black
      plays something as a computer, where the entropy of a black hole is determined by
      its event horizon area. That is to say, we can count the
      number of microstates of a black hole on a surface.
      Generally speaking, people should apply the quantum gravity  to
      the system where the density and curvature are large
      enough to the order of Planck scale, such as a black hole and the
      initial cosmic singularity. More or less wonderfully, in quantum field theory,
      the ultraviolet (UV) cut-off and
      infrared (IR) cut-off are suggested to be related to each other
      \cite{cohn}. Therefore, we should not be surprised if the
      dark energy problem is finally proved to be a
      problem  of quantum gravity, though its characteristic energy
      scale is very low.
      % A holographic dark energy is presented in \cite{li}, based on
      %the idea that the total energy in a volume should not larger
      %than the energy to form a black hole. But the black holes have
      %been formed everywhere in the universe.
      However, when we apply the holographic principle to the universe,
      the first problem we confront is which surface is the proper
      surface that the information of the volume ``holographying to"?
      In the case of black hole, the event horizon is a proper
      holography of the black hole. In the case of cosmology the
      result is not so evident. If we just take the event horizon as the holography
      for mimicking the case of black hole, we may be embarrassed in
      a decelerating universe, since it has no event horizon at all.
      So, we should find some more fundamental analogies between black
      hole physics and cosmology to find the proper surface.

      In the context of black hole physics we believe that the event horizon of a black hole
       is the proper surface since every physical quantity of the
       black hole, especially the entropy, shows itself properly on that
       surface. And they strictly obey the first law of the
       thermodynamics. An interesting progress is that Einstein equation can be  reproduced from the
       proportionality of entropy and horizon area together with
       the first law of thermal dynamics, $\delta Q = T
        dS$,  jointing to heat, entropy, and temperature, where the temperature is the Unruh
      temperature to an observer just behind a causal Rindler horizon
      \cite{jaco}. This work pioneers the way how to find the proper
      holography besides the case of black hole. In the case of
      dynamical solution, a similar procedure reproduces the Friedmann equation.
     One needs to apply only the first law of thermodynamics to the trapped surface (apparent horizon) of an FRW
     universe and assume the geometric entropy given
     by a quarter of the apparent horizon area and the temperature given by the inverse
     of the apparent horizon \cite{cai}. There are several arguments that the apparent horizon
     should   be a causal horizon and is associated with the gravitational
     entropy and Hawking temperature \cite{bak}. Hence it seems that the apparent
     horizon is the right holography of the universe \footnote{Note that in the case of a static black hole, the
     apparent horizon and the event horizon coincide each other. Thus we also can say the right
     holography of a black hole is the apparent horizon in that
     case.}.

     Up to now, our arguments are at the level of sophistication. Now we turn our sight to the realistic
     universe. There are several different components including baryon matter, non-baryon dark matter,
     and maybe dark energy etc, in our universe. We know the properties
     of baryon matter well. But, sadly, its entropy  obeys the ``volume
     law" rather than  the ``area law". And its entropy is far from
     saturating the holographic bound. Also the total entropy of all the known matters
     is still much lower than the holographic bound
     \cite{entropyu}. So it is interesting to see what will happen
     if a dark component of the universe strictly obeys the
     holographic principle?

     This paper is organized as follows: In the next section we will study a model in which a component
     satisfies the holographic principle. The dynamical analysis is left in section III. We find that there exists a
     stable dark matter-dark energy solution at late time, which is helpful to solve the coincidence problem.  We present our conclusion
     and some discussions in section IV.

  \section{the model }
   There are decisive evidences that our observable universe evolves adiabatically
   after inflation in a comoving volume, that is, there is no energy-momentum
   flow between different patches of the observable universe so that the universe keeps
   homogeneous and isotropic after inflation.  That is the reason why we can
   use an FRW geometry to describe the evolution  of the universe. In
   an  adiabatically evolving universe, the first law of
   thermodynamics equals the continuity equation. In
    a comoving volume the first law reads,
   \be
   dU=TdS-pdV,
   \label{1st}
   \en
   where $U=\Omega_k\rho a^3$ is the energy in this volume, $T$ denotes temperature,
   $S$ represents the entropy of this volume, and $V$ stands for the physical
   volume $V=\Omega_k a^3$. Here, $\Omega_k$ is a factor related to the spatial curvature,
   for spatially flat case $\Omega_0=\frac{4}{3}\pi$, in this paper we only consider the spatially flat model,
   $\rho$ is the energy density and $a$
   denotes the scale factor.
    For examples, in the case of radiation
   $p=1/3 \rho$, then we derive $\rho\sim a^{-4}$; in the case of
   dust $p=0$, then we obtain $\rho\sim a^{-3}$; and in the case
   of vacuum $p=-\rho$, then $\rho=$constant.

   Since we know little about the properties of
   dark energy, especially in the theoretical side, it is reasonable
   to study the possibility of a non-adiabatical dark energy.
   Next we consider the non-adiabatical evolutions. When the term $TdS$
   does not equal zero, the results are completely different. For
   instance, it is neither sufficient nor necessary that $\rho$=constant implies $p=-\rho$.
    We will show it is just
   the case if a dark component strictly obeys the holographic
   principle. To our knowledge, this point scarcely gets any attention in
   the literatures.

   Based on the investigations in \cite{bak, cai}, the entropy in
   the apparent horizon is
   \be
   S=\frac{8\pi^2 \mu^2}{H^2},
   \en
   where $H$ is the Hubble parameter, $\mu$ denotes the reduced
   Planck mass. This equation implies that the entropy is exactly one-fourth of
   the area of the apparent horizon.  So, in a comoving volume the entropy becomes,
   \be
   S_c=\frac{8\pi^2 \mu^2}{H^2} \frac{a^3}{H^{-3}}={8\pi^2
   \mu^2}Ha^3.
   \label{Sc}
   \en
    In the above equation we have used an assumption that
    the entropy is homogeneous in the observable universe.
    This is not a tough assumption for we have no good reason
    that the entropy density in one region is larger  than other
    regions.   Evidently, the entropy in a comoving volume is not constant.
    However, as we discussed before, our observable universe evolves adiabatically
    after inflation in any comoving volumes. Thus, if a dark
    component, which is called dark energy, satisfies the holographic principle, it requires the
    other compensative dark component, which is assumed to be dark matter, such that the total entropy in a
    comoving volume keeps  constant. In this sense, our universe
    only  partly obeys the holographic principle, which indicates
    the  title: Semi-Holographic Universe.

    With the above supposition and conventions the entropy of the dark energy satisfies (\ref{Sc}),
    \be
    S_{de}={8\pi^2
   \mu^2}Ha^3.
   \label{detro}
    \en
    Correspondingly, the entropy of dark matter in this comoving
    volume should be
    \be
    S_{dm}=C-S_{de},
    \en
   where $C$ is a constant, representing the total entropy of the
   comoving volume. The Hubble parameter $H$ is determined by the
   Friedmann equation,
   \be
   H^2=\frac{1}{3\mu^2}(\rho_{dm}+\rho_{de}+\Lambda),
   \label{fried}
   \en
   where $\rho_{dm}$ denotes the density of non-baryon dark matter,
   $\rho_{de}$ denotes the density of dark energy, and $\Lambda$ is
   the cosmological constant (or vacuum energy). In this preliminary
   research, we omit the baryon matter since its partition is very
   small and does little work in the late time universe. However we
   introduce the cosmological constant because we want  not only to
   consider a more general case, but also to show that the holographic
   dark energy and dark matter require each other even if there is a
   cosmological constant. And furthermore, we will see that the
   cosmological constant plays an important role in the final state
   of the universe.

  \section{dynamical analysis}
  To investigate the evolution in a more detailed way, we take a
  dynamical analysis of the universe.

   The holographic principle requires that the temperature
   \cite{cai}
   \be
   T=\frac{H}{2\pi}.
   \label{tem}
   \en
   By using (\ref{tem}), (\ref{fried}), and (\ref{detro}), the first
   law of thermal dynamics (\ref{1st}) becomes the evolution
   equation of dark energy,
   \be
   \frac{2}{3}\rho_{de} '=\rho_{dm}(1-w_{dm})-\rho_{de}(1+3w_{de})+2\Lambda,
   \label{evlde}
   \en
    where a prime denotes the derivative with respect to $\ln a$,
    $w_{dm}$ indicates the equation of state (EOS) of dark matter, and $w_{de}$ represents
    the EOS of dark energy. Similarly, we derive the evolution
    equation of dark matter,

   \be
   \frac{2}{3}\rho_{dm} '=-\rho_{dm}(3+w_{dm})+\rho_{de}(-1+w_{de})-2\Lambda.
   \label{evldm}
   \en
   For convenience we introduce two new dimensionless functions to
   represent the densities,
   \be
   u\triangleq \frac{\rho_{dm}}{3\mu^2H_0^2},
   \en
   \be
   v\triangleq \frac{\rho_{de}}{3\mu^2H_0^2},
   \en
  and a dimensionless cosmological constant
  \be
  \lambda\triangleq \frac{\Lambda}{3\mu^2H_0^2},
  \en
  where $H_0$ denotes the present Hubble parameter. Then the
  equation set (\ref{evldm}), (\ref{evlde}) becomes
  \be
  \frac{2}{3} u '=-u(3+w_{dm})+v(-1+w_{de})-2\lambda,
   \label{evldm1}
   \en
    \be
   \frac{2}{3}v '=u(1-w_{dm})-v(1+3w_{de})+2\lambda,
   \label{evlde1}
   \en
   respectively. We note that the  time variable does not appear in the dynamical system
  (\ref{evldm1}) and (\ref{evlde1}) because time has been completely
  replaced by scale factor.

  Before presenting the numerical examples for special parameters we
  study the analytical property of this system.
  The critical points of
  dynamical system (\ref{evldm1}) and (\ref{evlde1}) are given by
  \be
  u_c'=v_c'=0,
  \en
  which yields,
  \be
  u_c=-\lambda \frac{w_{\rm de}+1}{2w_{\rm de}+w_{\rm dm}w_{\rm de}+1},
  \en
  \be
  v_c=\lambda \frac{w_{\rm dm}+1}{2w_{\rm de}+w_{\rm dm}w_{\rm de}+1}.
  \en
  So, finally the universe enters a de Sitter phase, and the ratio of
  dark matter over dark energy is
  \be
  \frac{u_c}{v_c}=-\frac{1+w_{\rm de}}{1+w_{\rm dm}}.
  \en
   We see that the final ratio is independent of the cosmological
   constant. There are two reasonable cases: case I, $w_{\rm de}<-1$ and
   $w_{\rm dm}>-1$; case II, $w_{\rm de}>-1$ and
   $w_{\rm dm}<-1$, since we should require both of the final densities of dark matter and dark
   energy are positive.
   Physically, it is easy to understand. Since one of the dark
   components flows out entropy (surely with energy) to the other dark component, at the same
   time it keeps a constant density, its apparent EOS should be less than
   $-1$, like phantom, for similar mechanism, see \cite{self}.

   Now
   we consider the degenerated case in which $\lambda=0$. Under this
   condition the equation set (\ref{evldm1}) and (\ref{evlde1})
   become homogeneous. A non-trivial solution implies its determinate
   of coefficients equals zero,
   \be
   1+2w_{\rm de}+w_{\rm dm}w_{\rm de}=0.
   \en
   Under this condition the ratio of
  dark matter over dark energy reaches the same as in the case with a
  $\lambda$,
  \be
  \frac{u_c}{v_c}=-\frac{1+w_{\rm de}}{1+w_{\rm dm}}.
  \en
   But, this ratio of dark matter and dark energy will keep the
   same  value in the whole history of the universe, that is, dark matter and dark energy
   always evolve in the same way: It is not a
   very interesting case.

   If
   the present dark energy dominated universe can be an attractor of
   the dynamical evolution, it is helpful to overcome the
   coincidence problem. To confirm the critical point of the system is an attractor, we
   need the stability property of it. Imposing a perturbation
  to the critical points, we derive,
  \be
  \frac{2}{3}(\delta \rho_{dm})'=-\delta \rho_{dm}(3+w_{\rm dm})+\delta
  \rho_{de} (-1+w_{\rm de}),
  \en
  \be
  \frac{2}{3}(\delta \rho_{de})'=\delta \rho_{dm}(1-w_{\rm dm})-\delta
  \rho_{de} (1+3w_{\rm de}).
  \en
  Note that we have assumed both $w_{\rm dm}$ and $w_{\rm de}$ are constant from the beginning.
  The  eigenvalues of this system read,
  \be
  l_1=\frac{1}{2}\left(-w_{\rm dm}+w_{\rm de}+\sqrt{16+w_{\rm dm}^2+32w_{\rm de}+14w_{\rm dm}w_{\rm de}+w_{\rm de}^2}~\right),
  \en
   \be
  l_2=\frac{1}{2}\left(-w_{\rm dm}+w_{\rm de}-\sqrt{16+w_{\rm dm}^2+32w_{\rm de}+14w_{\rm dm}w_{\rm de}+w_{\rm de}^2}~\right),
  \en

  Stability implies all of real parts of the eigenvalues are less
  than zero, which requires,
  \be
  w_{\rm de}<{\rm Min}\left\{-1, -\frac{1}{2+w_{\rm dm}}\right\},
  \en
  when $w_{\rm dm}>-2$. The system will be unstable for any $w_{\rm de}$ when $w_{\rm dm}\leq
  -2$. Therefore, case I is stable while case II is unstable.

  The most significant parameter from the viewpoint of
  observations is the deceleration parameter $q$, which carries the total
  effects of cosmic fluids. Using (\ref{fried}), (\ref{evlde}), and
  (\ref{evldm}) we obtain the deceleration parameter in this model
  \be
  q=-\frac{\ddot{a}}{a} \frac{1}{H^2}
  =\frac{1}{2}\frac{\rho_{dm}(1+3w_{dm})+\rho_{de}(1+3w_{de})-2\lambda}
  {\rho_{dm}+\rho_{de}+\lambda}.
  \en
  For a numerical example, we take the terminal ratio of dark energy
  to dark matter $1:1$, the present dark matter partition
  $u_0=0.25$, the present holographic dark energy partition
  $v_0=0.01$, correspondingly $\lambda=0.74$.

  Figure 1 illuminates the evolution of deceleration parameter. As a
  simple example we just set $w_{dm}=-0.4$, $w_{de}=-1.4$. From the figure
  one see that current $q\sim -0.75$ and at the high redshift region
  it goes to $0.5$, which is well consistent with current observations \cite{review, WMAP}. As a comparison,
  we plot the evolution of the deceleration parameter in a spatially flat $\Lambda$CDM, in which we set $\Omega_{dm}=0.25$. One sees
  that the deceleration parameter more swiftly approaches $0.5$ (standard dark matter model, SCDM for short) in this semi-holographic model
  than that of $\Lambda$CDM.

  One may be confused why we can endow a negative EOS
  to the dark matter. In fact, its evolution do not depend on this
  apparent EOS, but the effective EOS. We define the effective EOS
  as the following procedure. Supposing the dark matter evolves adiabatically itself,
  we obtain its evolution from (\ref{1st}),
  \be
 {d\rho_{dm}}+3(\rho_{dm}+p_{eff})\frac{da}{a}=0,
 \label{em}
 \en
 where $p_{eff}$ denotes the effective pressure of dark matter. Then
 we obtain
 \be
 w_{dme}\triangleq \frac{p_{eff}}{\rho_{dm}}=\frac{1}{2}
 +\frac{1}{2}w_{\rm dm}+\frac{v}{2u}(1-w_{\rm de})+
 \frac{\lambda}{u},
 \en
 which is a variable in the evolution history of the universe.
 Figure 2 displays $w_{dme}$ as a function of $\ln a$, in which we set the same parameters
 as in figure 1. This figure
 shows that the dark matter is effectively very stiff in the current
 time, but quickly gets softer and becomes ordinary dust in a higher redshift region.
 The corresponding effective EOS of dark energy $w_{dee}$ is illuminated in Figure 3. From this figure we know that the  dark energy currently evolves as
 phantom and becomes a cosmological constant in some high redshift region.

 Associating figure 1 with figure 2 and 3, we conclude that the universe
 is dominated by dark matter, whose effective EOS $w_{eff}=0$, in
 some high redshift region, such as $z>1.5$ ($-\ln a>0.92$)and
 hence essentially SCDM recovers.

   The other point deserves to note is that we only introduce a little
   bit of holographic dark energy (in the above example 1\%), the final state changes
   heavily. It evolves into a dark energy-dark matter scaling
   solution, which shed light on the coincidence problem.
   The dark matter will be diluted rapidly if we do not introduce it, which yields coincidence.

  \begin{figure}
 \centering
 \includegraphics[totalheight=2.3in, angle=0]{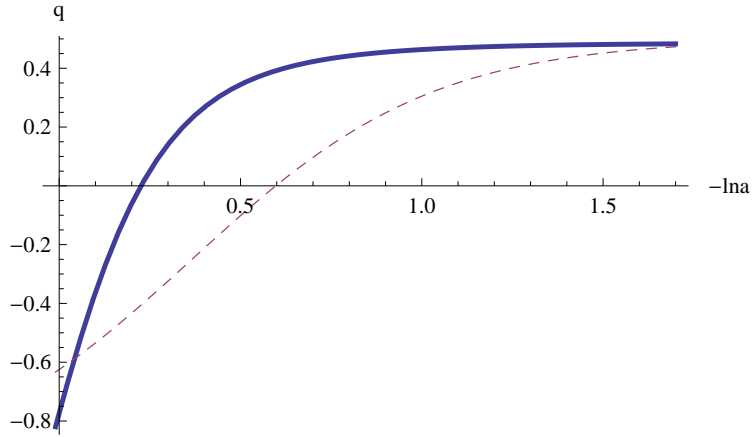}
 \caption{The evolutions of $q$ in semi-holographic universe (solid curve) and in $\Lambda$CDM (dashed curve), respectively.}
 \label{dece}
 \end{figure}

  \begin{figure}
 \centering
 \includegraphics[totalheight=2.3in, angle=0]{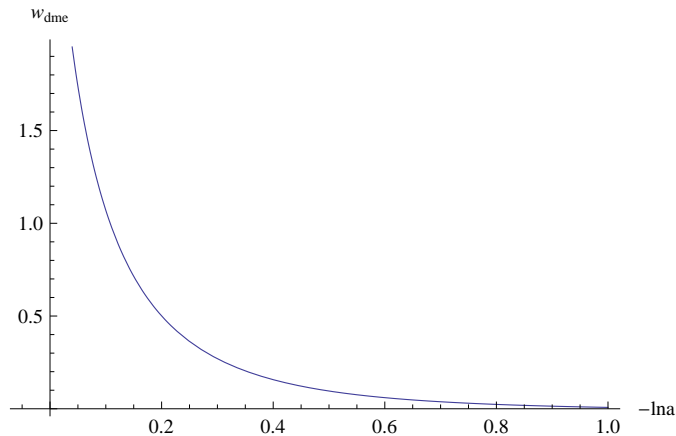}
 \caption{The effective EOS of dark matter $w_{dme}$ as a function of $-\ln a$.    }
 \label{weff}
 \end{figure}

 \begin{figure}
 \centering
 \includegraphics[totalheight=2.3in, angle=0]{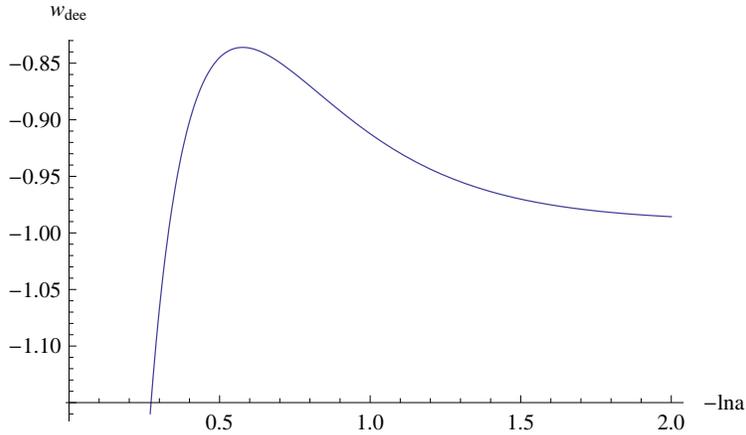}
 \caption{The effective EOS of dark energy $w_{dee}$ as a function of $-\ln a$.    }
 \label{weff}
 \end{figure}

%%%%%%%%%%%%%%%%%%%%%%%%%%%%%%%%%%%%%%%%%%%%%%%%%%%%%%%%%%%%%%%

\section{Conclusion and discussion}

 We present a cosmological model inspired by holographic principle, especially the previous studies of the
 relation between thermal dynamics and general relativity, in which the entropy of the dark
 energy is one-fourth of the area of the apparent horizon. We find that under this condition
 the dark energy must evolve non-adiabatically. But the total matter
 in a comoving volume should evolve adiabatically. Hence a
 compensating component should exist, which we called dark matter.

 We find a future attractor solution, which is a stable scaling solution for
 the dark matter-dark energy system in some proper region of the
 parameters $w_{\rm dm},~w_{\rm de}$. The final ratio of dark matter to dark energy
  only depends $w_{\rm dm},~w_{\rm de}$, which is independent on the initial
 values of the densities of dark matter and dark energy. This result
 is helpful to solve the coincidence problem.

 In a numerical example, we find that the deceleration parameter can
 be well consistent with observations.

 In this paper only a spatially flat universe is discussed. The
 model with a non-vanishing spatial curvature need to be
 investigated
 further. Also, as a phenomenological model, the parameters should be constrained by observation data further.

 The derivation of Fridmann equation from the assumption that the
 entropy is a quarter of its apparent horizon is strict and does not
 depend on the energy scale. Hence, we expect that the holographic
 component may have remarkable effects in the early universe.

 {\bf Acknowledgments.}
 H Zhang and X Li are supported by National Education Foundation of China under grant No. 200931271104,
 Shanghai Municipal Pujiang grant No. 10PJ1408100, and National Natural Science Foundation of China under Grant No. 11075106.  H.Noh is supported by grant No. C00022 from the Korea Research
 Foundation.

\end{document}